\documentclass[12pt,onecolumn]{IEEEtran}

\newcommand{\define}{\buildrel \Delta \over =}
\font\title=cmr17
\font\author=cmr12
\font\thanks=cmr7

\usepackage{amssymb}
\usepackage[final]{graphicx}
\usepackage{graphicx}
\usepackage{pslatex}
\usepackage{epstopdf}
\usepackage{subfig}

\begin{document}

\begin{titlepage}

\begin{center}
\huge A Universal Grammar-Based Code\end{center}
\begin{center}\huge For Lossless Compression of Binary Trees
\end{center}

\bigskip
\vskip 0.5in
\centerline{\author Jie Zhang$^*$,
        En-hui Yang$^{**}$ and John C.\ Kieffer$^{***}$}


\vskip 5.1truein
\vrule height .01 in width 2.5 in \par

{\thanks \noindent$^*$Avoca Technologies Inc., 563 Edward Ave., Suite 13, Richmond Hill, Ontario CA.}\newline
{\thanks \noindent$^{**}$Dept.\ of Electrical \& Computer Engineering, University of Waterloo, Waterloo, Ontario CA.}
\newline
{\thanks $^{***}$Dept.\ of Electrical \& Computer Engineering, University of Minnesota Twin Cities, Minneapolis, Minnesota USA.} \par

\par

\end{titlepage}

\begin{abstract}
We consider the problem of lossless compression of binary trees, with the aim of reducing
the number of code bits needed to store or transmit such trees.
  A lossless grammar-based code is presented which encodes each binary tree into a
  binary codeword in two steps. In the first step, the tree
is transformed into a context-free grammar from which the tree can be reconstructed. In the second step,  the context-free grammar is encoded
into a binary codeword. The decoder of the grammar-based code decodes the original tree from
 its codeword by reversing the two encoding steps.  It is shown that the resulting grammar-based binary tree compression code is a universal code on a family of
 probabilistic binary tree source models satisfying certain weak restrictions.
\end{abstract}


\section{Introduction}

There have been some recent initial attempts to conceptualize the notion of structure in information theory \cite{isit2009}\cite{choi}\cite{isit2010},
with the ultimate future goal being the development of a lossless compression theory for structures. In the present paper, we put
forth a general framework for this area, and then develop a lossless compression theory
for binary tree structures within this framework.
Our framework will permit an abstract asymptotic theory for the compression of structures
to be developed, where the framework is sufficiently general to include the types of structures that have been considered
in other contexts, such as in the asymptotic theory of networks \cite{lov} or the asymptotic theory of patterns \cite{grenander}.
The basic concepts in this framework are the notions of {\it structure universe}, {\it structure filter}, and {\it structure source},
which we now define; after the definitions, we give examples of the concepts relevant for the work we shall do in this paper.
\par
{\it Concept of Structure Universe.} Broadly speaking, ``structure universe'' will mean the set of structures
under consideration in a particular context. Each structure has a ``size'' assigned to it, which is a positive integer that can be
a measure of
how large or how complex the structure is. For example, if a structure is a finite graph $g$, then the size of the structure could
be taken as the number of vertices of $g$ or the number of edges of $g$; if a structure is a finite tree $t$, then the size
of the structure could be taken as the number of leaves of $t$. We now make the notion of structure universe precise.
A {\it structure universe} $\Omega$ is defined to be any countably infinite set such that for each $\omega\in\Omega$ there
is defined a positive integer $|\omega|$, which we call the size of $\omega$,  such that the set
$\{\omega\in\Omega:|\omega|=n\}$ is finite for each positive integer $n$.

\par
{\it Concept of Structure Filter.} A structure filter $\cal F$
over a structure universe $\Omega$ (called  $\Omega$-filter for short) is defined to be any
set of finite nonempty subsets of $\Omega$ which forms a partition of $\Omega$.
For example, given any structure universe $\Omega$, we have the natural $\Omega$-filter
consisting of all nonempty subsets of $\Omega$ of the form $\{\omega\in\Omega: |\omega|=n\}$
($n=1,2,\cdots$).
Given an $\Omega$-filter $\cal F$, a real-valued function $(x_F:F\in{\cal F})$ defined on $\cal F$, and an extended real number $L$,
the limit statement
$\lim_{F\in {\cal F}}x_F=L$ means that for any neighborhood $\cal N$ of $L$ in  the topology of the extended real line,
the set $\{F\in {\cal F}:x_F \not \in {\cal N}\}$ is finite; the limit $L$, if it exists, is unique, which is due to the fact
that a structure filter is always countably infinite. Similarly, one can make sense of limit statements of the
form $\limsup_{F\in{\cal F}}x_F=L$ and $\liminf_{F\in{\cal F}}x_F=L$.
The sets in any $\Omega$-filter $\cal F$ are growing in the sense that
\begin{equation}
\lim_{F\in{\cal F}}\left[\min\{|\omega|: \omega \in F\}\right]=\infty.
\label{mar28eq1}\end{equation}
This condition will make possible an asymptotic theory of lossless compression of structures; we will see how the condition is used in Sec.\ III.

\par
{\it Concept of Structure Source.} Informally, suppose we randomly select
a structure from each element of a structure filter; then these random structures constitute the output of a structure source. Formally,
we define a structure source to be any triple $(\Omega,{\cal F}, P)$ in which $\Omega$ is a structure universe,
$\cal F$ is an $\Omega$-filter, and $P$ is a function from $\Omega$ into $[0,1]$ such that
\begin{equation}
\sum_{\omega\in F}P(\omega) = 1,\;\;F \in {\cal F}.
\label{mar28eq2}\end{equation}
Note that (\ref{mar28eq2}) simply tells us that $P$ restricted to each $F \in {\cal F}$ yields a probability distribution on $F$;
for any subset $F'$ of $F$, we write the probability of $F'$ under this distribution as $P(F')$, which is computed as the
sum $\sum_{\omega\in F'}P(\omega)$.
\par

{\it Example 1.} For each $n\geq 2$, fix an undirected graph $g_n$
with $n$ vertices and $n(n-1)/2$ edges, one edge for each pair of distinct vertices, and let $G_n$ be
the set of edge-labelings of $g_n$ in which each edge of $g_n$ is assigned a label from the set $\{0,1\}$. That is, $G_n$
consists of all pairs $(g_n,\alpha)$ in which $\alpha$ is a mapping from the set
of edges of $g_n$ into the set $\{0,1\}$.  Let ${G}_n^*$ be a subset of ${G}_n$ such that for each $(g_n,\alpha) \in { G}_n$, there exists a unique
$(g_n,\alpha^*)\in { G}_n^*$ into which $(g_n,\alpha)$ is carried by an isomorphism (that is,
there is an isomorphism of $g_n$ onto itself which carries each edge $e$
of $g_n$ into an edge $e'$ of $g_n$ for which the edge labels $\alpha(e),\alpha^*(e')$ coincide).
For example, $G_3^*$ consists of four edge labelings of $g_3$, one in which all three of the edges of $g_3$ are labeled $0$, a second one in which all edge labels
are $1$, a third one in which two edge labels are $0$ and the remaining one is $1$, and a fourth one in which two  edge labels are $1$
and the remaining one is $0$.
Let $\Omega$ be the structure universe $\cup_{n\geq 2}G_n^*$, where we define the size of
each labeled graph in $\Omega$ to be the number of vertices of the graph.
Let $\cal F$ be the $\Omega$-filter $\{G_n^*:n\geq 2\}$.
For each $\sigma\in (0,1)$, let $S_{\sigma}=(\Omega,{\cal F},P_{\sigma})$ be the structure source
such that for each $(g_n,\alpha') \in \Omega$,
$$P_{\sigma}(g_n,\alpha')= N(g_n,\alpha')\sigma^{m_1}(1-\sigma)^{m_0},$$
where $m_0$ is the number of edges of $g_n$ assigned $\alpha'$-label $0$,
$m_1$ is the number of edges of $g_n$ assigned $\alpha'$-label $1$,
and $N(g_n,\alpha')$ is the number of $(g_n,\alpha)$ belonging to $G_n$ for which
$(g_n,\alpha^*)=(g_n,\alpha')$. For example, the $P_{\sigma}$ probabilities assigned to
the four structures in $G_3^*$ given above are $\sigma^3$,$(1-\sigma)^3$, $3\sigma^2(1-\sigma)$, and $3\sigma(1-\sigma)^2$,
respectively.
In random graph theory, the structure source ${S}_{\sigma}$ is called the Gilbert model \cite{gilbert}.
 Choi and Szpankowski \cite{choi}
addressed the universal coding problem for the parametric family of sources  $\{{S}_{\sigma}: 0 < \sigma < 1\}$.
(We discuss universal coding  for general structure sources after the next two examples.)
\par

{\it Example 2.} We consider finite rooted binary trees having
at least two leaves such that each non-leaf vertex has exactly two ordered children.
From now on,
the terminology ``binary tree'' without further qualification will automatically
mean such a tree.
Let $\cal T$ be a set of binary trees such that each binary tree is isomorphic as an ordered tree
to a unique tree in $\cal T$. Then $\cal T$ is a structure universe, where
the size $|t|$ of a tree $t$ in the universe $\cal T$ is taken to be the number of leaves
of $t$.
We discuss two ways in which $\cal T$ can be partitioned to obtain a $\cal T$-filter.
For each $n\geq 2$,
let ${\cal T}_n$ be the set of trees in $\cal T$ that have $n$ leaves.
For each $n\geq 1$, let ${\cal T}^n$ be the set of trees in $\cal T$ for which the longest root-to-leaf
path consists of $n$ edges (that is, ${\cal T}^n$ consists of trees of depth $n$).
Then ${\cal F}_1=\{{\cal T}_n:n\geq 2\}$ and ${\cal F}_2=\{{\cal T}^n:n\geq 1\}$ are each ${\cal T}$-filters.
A structure source of the form $({\cal T},{\cal F},P)$ for some ${\cal T}$-filter $\cal F$ is called a {\it binary tree source}.
In \cite{isit2009}, binary tree sources of form $({\cal T},{\cal F}_1,P)$ were
introduced which are called leaf-centric binary tree source models; we address the universal coding problem for
such sources in Section IV of the present paper.
 In Section V, we address the universal coding problem for a type of binary tree source of form $({\cal T},{\cal F}_2,P)$ which
we call a depth-centric binary tree source model.
\par

 {\it Example 3.} Let $A$ be a finite alphabet. For each $n\geq 1$, let $A^n$ be the set of
 all $n$-tuples of entries from $A$. Then $\Omega=\cup_{n=1}^{\infty}A^n$ is a structure universe,
 where we define the size of each structure in $A^n$ to be $n$.
 Let ${\cal F}$ be the $\Omega$-filter $\{A^n:n\geq 1\}$. A structure source
 of the form $(\Omega,{\cal F},P)$ corresponds to the classical notion of finite-alphabet
 information source (\cite{han}, page 14) . Thus, source coding theory for structure sources will include
 classical finite-alphabet source coding theory as a special case.
 \par

{\it Asymptotically Optimal Codes for Structure Sources.}  In the following and in the rest of the paper, $\cal B$ denotes the set of non-empty finite-length
binary strings, and $L[b]$ denotes the length of string $b \in {\cal B}$. Let $\Omega$ be a structure universe. A lossless code on $\Omega$ is a pair $(\psi_e,\psi_d)$ in which
\begin{itemize}
\item $\psi_e$ (called the encoding map) is a one-to-one mapping of $\Omega$ into $\cal B$
 which obeys the prefix condition, that is, if $\omega_1$ and $\omega_2$
 are two distinct structures in $\Omega$, then $\psi_e(\omega_1)$ is not a prefix of $\psi_e(\omega_2)$; and
 \item $\psi_d$ (called the decoding map) is the mapping from $\psi_e(\Omega)$ onto $\Omega$
which is the inverse of $\psi_e$.
\end{itemize}
Given a lossless code $(\psi_e,\psi_d)$ on structure universe $\Omega$ and a structure source $(\Omega,{\cal F},P)$,
then for each $F \in {\cal F}$ we define the real number
$$R(\psi_e,F,P) \define \sum_{\omega\in F,\;P(\omega)>0}|\omega|^{-1}\{L[\phi_e(\omega)]+\log_2P(\omega)\}P(\omega),$$
which is called the $F$-th order average redundancy of the code $(\psi_e,\psi_d)$ with respect to the source.
We say that a lossless code $({\psi}_e,{\psi}_d)$ on $\Omega$ is an asymptotically optimal
code for structure source $(\Omega,{\cal F},P)$ if
\begin{equation}
\lim_{F\in{\cal F}}R({\psi}_e,F,P) = 0.
\label{apr5eq1}\end{equation}
\par

{\it Universal Codes for Structure Source Families.} Let $\cal F$ be a fixed $\Omega$-filter for structure universe $\Omega$.  Let $\mathbb{P}$ be a set of mappings from  $\Omega$ into $[0,1]$ such that
(\ref{mar28eq2}) holds for every $P\in{\mathbb{P}}$. A universal code for the family of structure sources $\{(\Omega,{\cal F},P):P\in{\mathbb{P}}\}$   (if it exists) is a lossless code
on $\Omega$ which is asymptotically optimal for every source in the family.
The universal source coding problem for a family of structure sources is to determine whether the family
has a universal code, and, if so, specify a particular universal code for the family.
\par

There has been little previous work on universal coding of structure sources. One notable exception is the
work of Choi and Szpankowski \cite{choi}, who devised a universal code for the parametric family
of Gilbert sources $\{{S}_{\sigma}:0< \sigma <1\}$ introduced in Ex.\ 1. Peshkin \cite{pesh}
and Busatto et al. \cite{maneth2} proposed grammar-based codes  for compression of general graphical structures and binary tree structures, respectively; as these authors did not
use a probabilistic structure source model, it is unclear whether their codes are
universal in the sense of the present paper (instead, they tested performance of their codes
on actual structures).\par

{\it Context-Free Grammar Background.} In the present paper, we further develop the idea
behind the Busatto et al. code \cite{maneth2} to obtain a grammar-based code for binary trees
which, under weak conditions, we prove to be a universal
code for families of binary tree sources. In this Introduction, we describe the
structure of our code in general terms; code implementation details will be given in Section II.
In order to describe the grammar-based nature of our code, we need at this point to give
some background information concerning deterministic context-free grammars. A deterministic context free grammar
${\mathbb{G}}$ is a quadruple $(S_1,S_2,s^*,P)$ in which
\begin{itemize}
\item $S_1$ is a finite nonempty set whose elements are called the nonterminal variables
of ${\mathbb{G}}$.
\item $S_2$ is a finite nonempty set whose elements are called the terminal variables
of ${\mathbb{G}}$. ($S_1\cup S_2$ is the complete set of variables of ${\mathbb{G}}$.)
\item $s^*$ is a designated nonterminal variable called the start variable of ${\mathbb{G}}$;
\item $P$ is the finite set of production rules of production rules of ${\mathbb{G}}$. $P$
has the same cardinality as $S_1$. There is
exactly one production rule for each nonterminal variable $s$, which takes the form
\begin{equation}
s\to (s_1,s_2,\cdots,s_n),
\label{apr19eq1}\end{equation} where $n$ is a positive integer which can depend on the rule
and $s_1,s_2,\cdots,s_n$ are variables of ${\mathbb{G}}$. $s$, $(s_1,\cdots,s_n)$, and $n$ are
respectively called the left member, right member, and arity of the rule (\ref{apr19eq1}).
\end{itemize}
Given a deterministic context-free grammar ${\mathbb{G}}$, there is a unique up to isomorphism rooted ordered vertex-labeled tree
$t({\mathbb{G}})$ (which can be finite or infinite) satisfying the following properties:
\begin{itemize}
\item The label on the root vertex of $t({\mathbb{G}})$ is the start variable of ${\mathbb{G}}$.
\item The label on each non-leaf vertex of $t({\mathbb{G}})$ is a nonterminal variable of ${\mathbb{G}}$.
\item The label on each leaf vertex of $t({\mathbb{G}})$ is a terminal variable of ${\mathbb{G}}$.
\item Let $s(v)$ be the variable of ${\mathbb{G}}$ which is the label on each vertex $v$ of $t({\mathbb{G}})$.  For each non-leaf vertex
$v$ of $t({\mathbb{G}})$ and its ordered children $v_1,v_2,\cdots,v_n$,
$$s(v) \to (s(v_1),s(v_2),\cdots,s(v_n))$$
is a production rule of ${\mathbb{G}}$.
\end{itemize}
``Unique up to isomorphism'' means that for any two such rooted ordered trees there is an isomorphism between the trees as ordered
trees  that preserves
the labeling (that is, corresponding vertices under the isomorphism have the same label). We call $t({\mathbb{G}})$ the derivation tree of ${\mathbb{G}}$.
\par
{\it Outline of Binary Tree Compression Code.} Let $\cal T$ be the structure universe of binary trees introduced in Ex.\ 2.
Suppose $t\in{\cal T}$ and suppose ${\mathbb{G}}$ is a deterministic context-free grammar
such that the arity of each production rule is two.
Then we say that ${\mathbb{G}}$ forms a representation of  $t$ if
$t$ is the unique tree in $\cal T$ isomorphic as an ordered tree to the tree which results when all
vertex labels on the derivation tree of ${\mathbb{G}}_t$ are removed.
In Section II, we will assign to each $t\in {\cal T}$ a particular deterministic context-free grammar ${\mathbb{G}}_t$ which forms a representation of $t$.
Then we will assign to ${\mathbb{G}}_t$ a binary codeword $B({\mathbb{G}}_t)$ so that the prefix condition is satisfied.
The grammar-based binary tree code of this paper is then the lossless code $(\phi_e,\phi_d)$ on $\cal T$ in which the encoding map
$\phi_e$ and decoding map $\phi_d$ each operate in two steps as follows.
\begin{itemize}
\item {\bf Encoding Step 1:} Given binary tree $t\in{\cal T}$, obtain the context-free grammar ${\mathbb{G}}_t$ from $t$.
\item {\bf Encoding Step 2:} Assign to grammar ${\mathbb{G}}_t$ the binary word $B({\mathbb{G}}_t)\in {\cal B}$, and then $B({\mathbb{G}}_t)$ is the
codeword $\phi_e(t)$ for $t$.
\item {\bf Decoding Step 1:} The grammar ${\mathbb{G}}_t$ is obtained from $B({\mathbb{G}}_t)$, which is the inverse of the second encoding step.
\item {\bf Decoding Step 2:} ${\mathbb{G}}_t$ is used to obtain the derivation tree of ${\mathbb{G}}_t$, from which $t$ is obtained by removing all labels.
\end{itemize}
The two-step encoding/decoding maps $\phi_e$ and $\phi_d$ are depicted schematically in the following diagrams:
$$\fbox{Encoding Map ${\phi_e:\;\; t\in{\cal T}\;\;{\buildrel {\rm 1st}\;{\rm step}\over \longrightarrow}\;\; {\mathbb{G}}_t\;\; {\buildrel {\rm 2nd}\;{\rm step}\over \longrightarrow}\;\; B({\mathbb{G}}_t)=\phi_e(t)\in{\cal B}}$}$$
$$\fbox{Decoding Map $\phi_d:\;\;B({\mathbb{G}}_t)\;\;{\buildrel {\rm 1st}\;{\rm step}\over \longrightarrow}\;\; {\mathbb{G}}_t\;\; {\buildrel {\rm 2nd}\;{\rm step}\over \longrightarrow}\;\; t=\phi_d(B({\mathbb{G}}_t))$}$$

\par

We point out the parallel between the grammar-based binary tree compression algorithm of this paper and
the grammar-based lossless data compression methodology
for data strings presented in
\cite{Kieffer-Yang1}. In the grammar-based approach to compression of a data string $x$,
one transforms $x$ into a deterministic context-free grammar ${\mathbb{G}}_x$ from which $x$ is uniquely recoverable
as the sequence of labels on the leaves of the derivation tree of ${\mathbb{G}}_x$; one then compresses ${\mathbb{G}}_x$ instead of $x$ itself.
Similarly, in the grammar-based approach to binary tree compression presented here, one transforms
a binary  tree $t$ into the deterministic context-free grammar ${\mathbb{G}}_t$ from which $t$ is uniquely recoverable
by stripping all labels from the derivation tree of ${\mathbb{G}}_t$; one then compresses ${\mathbb{G}}_t$
 instead of $t$ itself.

\par
The rest of the paper is laid out as follows. In Sec.\ II, we present the implementation details of the grammar-based
binary tree compression code $(\phi_e,\phi_d)$.
In Sec.\ III, we present some weak conditions on a binary tree source
under which $(\phi_e,\phi_d)$ will be an asymptotically optimal code for the source.
The remaining sections exploit these conditions to arrive at
wide families of binary tree sources
on which $(\phi_e,\phi_d)$ is a universal code (families of leaf-centric
models in Sec.\ IV and families of depth-centric models in Sec.\ V).

\section{Implementation of Binary Tree Compression Code}

This section is organized as follows. In Section II-A, we give some background regarding binary trees that shall
be used in the rest of the paper. Then, in Sec II-B, we explain how to transform each binary tree $t\in{\cal T}$ into the
deterministic context-free grammar ${\mathbb{G}}_t$; this is Step 1 of encoding map $\phi_e$. In Section II-C, there follows an explanation on how the codeword $B({\mathbb{G}}_t)$
is obtained from ${\mathbb{G}}_t$; this is Step 2 of encoding map $\phi_e$. Examples illustrating the workings of  the encoding map $\phi_e$ and the decoding map $\phi_d$
are presented in Section II-D. Theorem 1 is then presented in Section II-E, which
gives a performance bound for the code $(\phi_e,\phi_d)$. Finally, in Section II-F, we discuss
 a sense in which the grammar ${\mathbb{G}}_t$ is minimal and unique among all grammars which form a representation of $t \in{\cal T}$.
\par

\subsection{Binary Tree Background}

 We take the direction along each
edge of a binary tree to be away from the root.
The root vertex of a binary tree is the unique vertex which
is not the child of any other vertex, the leaf vertices are the vertices that
have no child, and each of the non-leaf vertices has exactly two ordered children.
We regard a tree consisting of just one vertex to be a binary tree, which
we call a trivial binary tree; all other binary trees have at least two leaves and are called non-trivial.
Given a binary tree $t$,
$V(t)$ shall denote the set of its vertices, and $V^1(t)$ shall denote
the set of its non-leaf vertices.
Each edge of $t$ is an ordered pair $(a,b)$ of vertices in $V(t)$, where
$a$ is the vertex at which the edge begins and $b$ is the vertex at which the edge ends ($a$ is the parent of $b$ and $b$ is  a child of $a$).
A path in a binary tree is defined to be any sequence $(v_1,v_2,\cdots,v_k)$ of vertices
of length $k \geq 2$ in which each vertex from $v_2$ onward
is a child of the preceding vertex. For each vertex $v$ of a binary tree which is not the root, there is a unique path
which starts at the root and ends at $v$. We define the depth level of each non-root vertex $v$ of a binary tree to be one less
than the number of vertices in the unique path from root to $v$ (this is the number
of edges along the path); we define the depth level of the root to
be zero.
Vertex $v_2$ is said to be a descendant of vertex $v_1$
if there exists a (necessarily unique) path leading from $v_1$ to $v_2$.
If a binary tree has $n$ leaf vertices, then it has $n-1$ non-leaf vertices
and therefore $2(n-1)$ edges.
\par

We have a locally defined order on each binary tree $t$
in which each sibling pair of child vertices of $t$ is ordered.
 From this locally defined order, one can infer
various total orders on $V(t)$ which are each consistent
with the local orders on the sets of children. The most useful of the
possible total orders for us will be the {\it breadth-first} order.
If we list the vertices of a binary tree in breadth-first order, we
 first list the root vertex at depth level $0$, then its two ordered children at depth level $1$,
 then the vertices at depth level $2$, depth level $3$, etc.
Two vertices $v_1,v_2$ at depth level $j>0$ are consecutive in breadth-first order
if and only if either (a) $v_1,v_2$ have the same parent and $v_1$ precedes $v_2$
in the local ordering of children, or (b) the parent of $v_1$ and the parent
of $v_2$ are consecutive in the breadth-first ordering of the non-leaf vertices at depth level $j-1$.
It is sometimes convenient to represent a tree $t$ pictorially
via a  ``top down'' picture, where
the root vertex of $t$ appears at the top of the picture (depth level $0$) and edges extend downward in the picture
to reach vertices of   increasing depth level; the vertices at each depth level
will appear horizontally in the picture with their left-right order corresponding to
the breadth-first order. Fig.\ 1 depicts two binary trees with their vertices
labeled in breadth-first order.
\par

The structure universe $\cal T$ consists only of nontrivial binary trees. Sometimes we need to consider a trivial binary tree consisting of just one vertex.
Fix such a trivial tree $t^*$. Then ${\cal T}^*={\cal T}\cup\{t^*\}$ can be taken
as our structure universe of binary trees both trivial and nontrivial.
For each $n\geq 1$, letting ${\cal T}_n$ be the set of trees in ${\cal T}^*$ having $n$ leaves,
and letting $K_n$ be the cardinality of ${\cal T}_n$,
it is well known \cite{stanley} that
$\{K_n:n\geq 1\}$ is the Catalan sequence, expressible by the formula
$$K_n = \frac{1}{n}{2(n-1)\choose n-1},\;\;n\geq 1.$$
For example, using this formula, we have
$$K_1=K_2=1,\;\;K_3=2,\;\;K_4=5,\;\;K_5=14.$$
Fig.\ 1 depicts one of the $(1/8){14\choose 7}=429$ binary trees
in ${\cal T}_8$, and
one of the $(1/16){30\choose 15}=9,694,845$ binary trees in ${\cal T}_{16}$.
\par

\begin{figure}[htp]
\centering
\fbox{\includegraphics[scale=0.5]{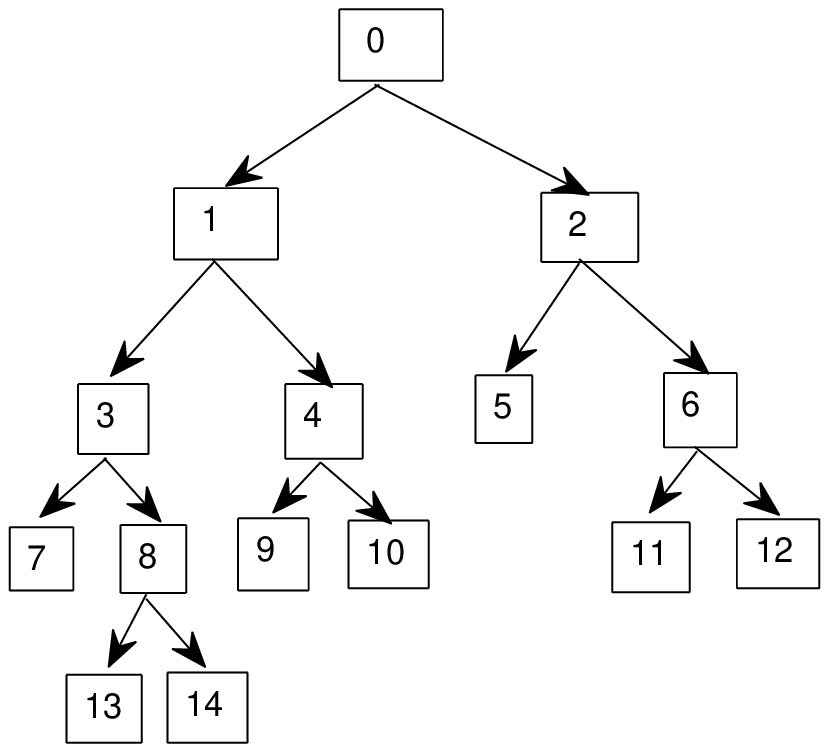}}
\fbox{\includegraphics[scale=0.5]{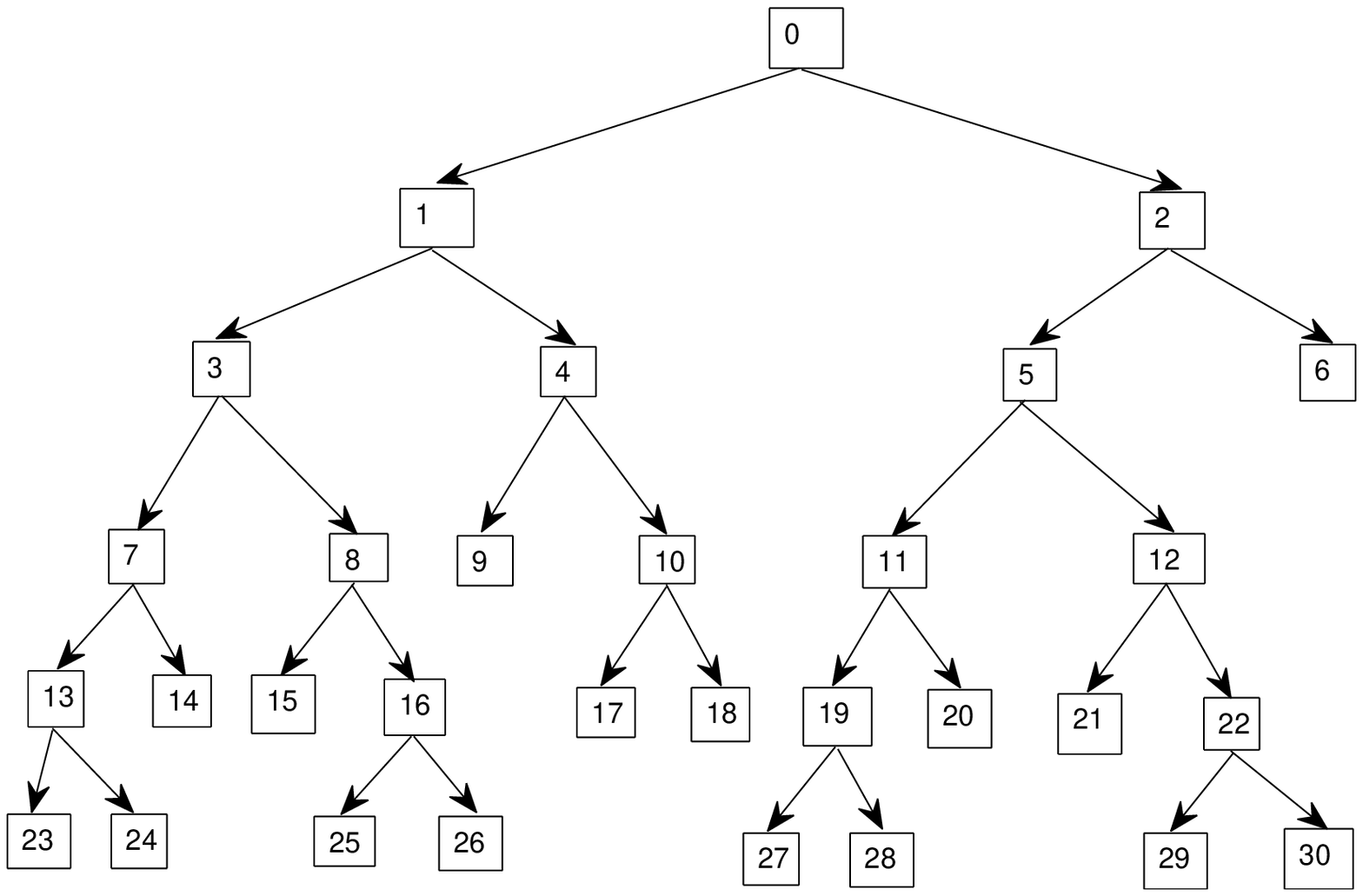}}
\caption{Binary trees in ${\cal T}_8$ (left) and ${\cal T}_{16}$ (right) with breadth-first ordered vertices}
\end{figure}

A subtree of a binary tree $t$ is a tree whose edges and vertices are edges and vertices of $t$; by convention, we require
also that a subtree of a binary tree should be a (nontrivial or trivial) binary tree. There
are two special types of subtrees of a binary tree  that shall be of interest to us, namely
final subtrees and initial subtrees. Given a binary tree $t$,
 a final subtree of $t$
is a subtree of $t$ whose root is some fixed vertex of $t$
and whose remaining vertices are all the descendants of this fixed vertex in $t$;
an initial subtree of $t$ is any subtree of $t$ whose root coincides with the root of $t$.
If $t$ is any nontrivial binary tree and $v\in V(t)$, we define $t(v)$ to be the unique binary tree
in ${\cal T}^*$ which is isomorphic to the final subtree of $t$ rooted at $v$.
Note that $t(v)=t^*$ if $v$ is a leaf of $t$, and that $t(v)=t$ if $t\in {\cal T}$ and $v$ is the root of $t$.
There are also two other trees of the $t(v)$ type which appear often enough that we give them a special name;
letting $v_1,v_2$ be the ordered children of the root of nontrivial binary tree $t$, we define $t_L=t(v_1)$
and $t_R=t(v_2)$ to reflect the respective left and right positions of these trees in the top down pictorial representation of tree $t$.

\subsection{Encoding Step 1}

Given $t\in{\cal T}$, we explain how to transform $t$ into the grammar ${\mathbb{G}}_t$, which is
Step 1 of the encoding map $\phi_e$. Define $N=N(t)$ to be the cardinality
of the set $\{t(v):v\in V(t)\}$. Note that $N\geq 2$ since $t^*$ and $t$ are distinct
and both belong to this set.
The set of nonterminal variables of  ${\mathbb{G}}_t$ is the nonempty set of integers $\{0,1,\cdots,N-2\}$.
The set of terminal variables of ${\mathbb{G}}_t$ is the singleton set $\{T\}$, where we have denoted the unique terminal variable
as the special symbol $T$. The start variable of ${\mathbb{G}}_t$ is $0$. All that remains to complete the definition
of ${\mathbb{G}}_t$ is to specify the production rules of ${\mathbb{G}}_t$. We do this indirectly by first labeling the vertices
of $t$ in a certain way and then extracting the production rules from the labeled tree. This labeling takes place as follows.
The root of $t$ is labeled $0$ and each leaf of $t$ is labeled ${T}$. The vertices of $t$ are traversed in breadth-first order. Whenever a vertex $v$ is thus
encountered which as yet has no label, one checks to see whether  $t(v)$ coincides  with $t(v')$ for some previously
traversed vertex $v'$. If this is the case, $v$ is assigned the same label as $v'$; otherwise, $v$ is assigned label equal to
the smallest member of the set $\{0,1,\cdots,N-2\}$ which has so far not been used as a label. For each nonterminal variable
$i \in \{0,1,\cdots,N-2\}$, we can then extract from the labeled tree the unique production rule of ${\mathbb{G}}_t$
of form $i\to (i_1,i_2)$ by finding any vertex of the labeled tree whose label is $i$; the entries $i_1,i_2$ are then
the respective labels on the ordered children of this vertex. Incidentally, the labeled tree we employed in this
construction turns out to be the derivation tree of ${\mathbb{G}}_t$.

\par
Figures 2-3 illustrate the results of Encoding Step 1 for the binary trees in Fig.\ 1.

\par

\begin{figure}[ht]
    \begin{minipage}[b]{0.48\linewidth}
     \centering
        \includegraphics[scale=0.5]{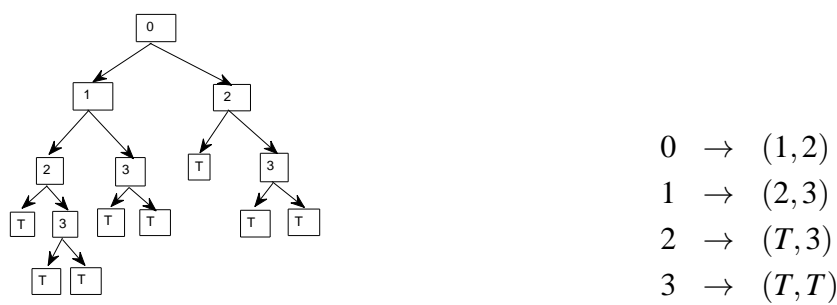}
    \end{minipage}
    \begin{minipage}[b]{0.48\linewidth}
      \begin{eqnarray*}
0 &\to & (1,2)\\
1 &\to & (2,3)\\
2 &\to& ({ T},3)\\
3 &\to& ({ T},{ T})
\end{eqnarray*}
\end{minipage}
    \caption{Encoding Step 1 For Left Figure 1 tree}
\end{figure}

\begin{figure}[ht]
    \begin{minipage}[b]{0.48\linewidth}
     \centering
        \includegraphics[scale=0.5]{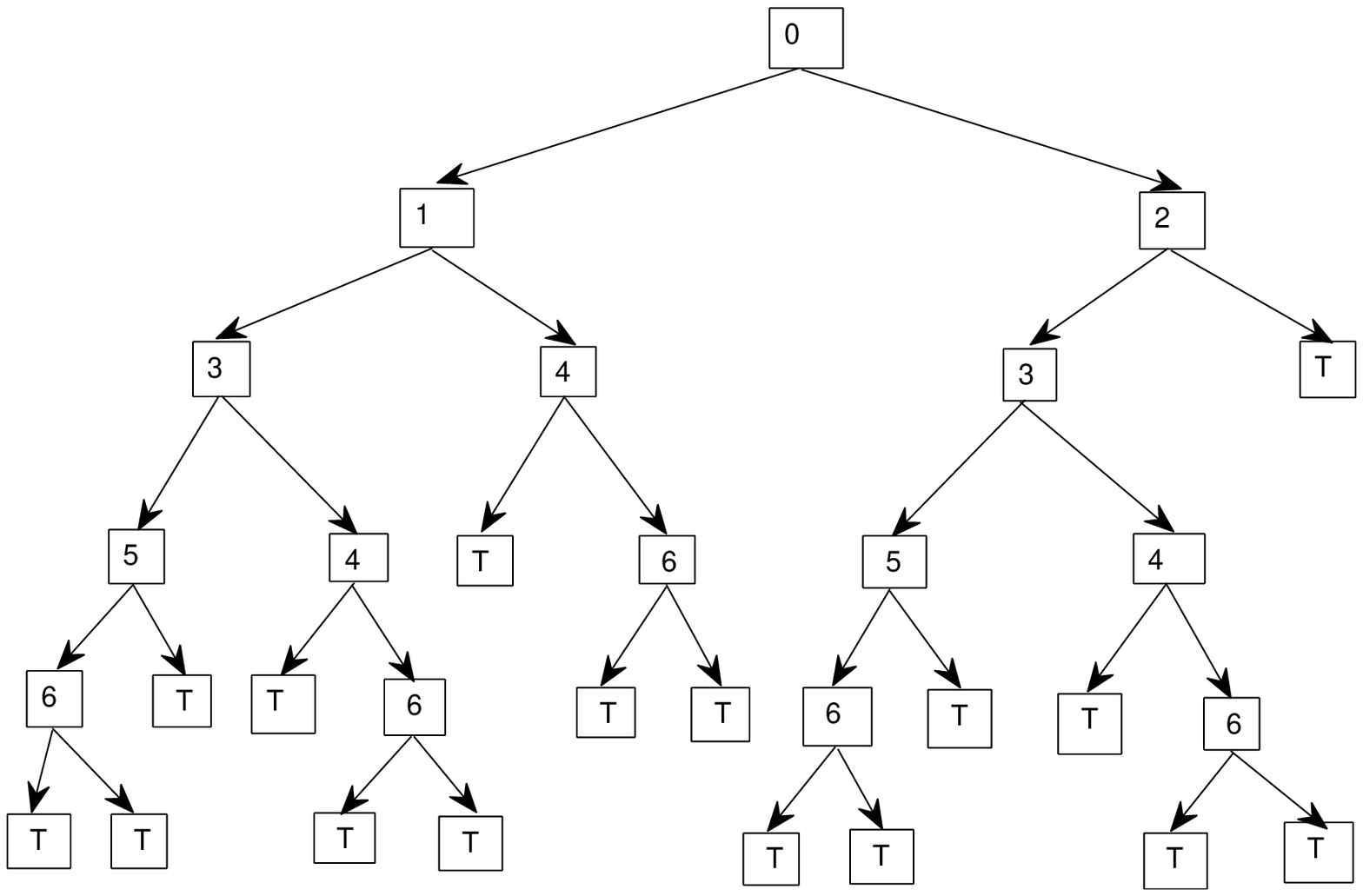}
    \end{minipage}
    \begin{minipage}[b]{0.48\linewidth}
      \begin{eqnarray*}
0 &\to& (1,2) \\
1 &\to& (3,4)\\
2 &\to& (3,{ T}) \\
3 &\to& (5,4)\\
4 &\to& ({ T},6) \\
5 &\to& (6,{ T}) \\
6 &\to& ({ T},{ T})
\end{eqnarray*}
     \end{minipage}
    \caption{Encoding Step 1 For Right Figure 1 tree}
\end{figure}

\subsection{Encoding Step 2}

Fix $t \in {\cal T}$. We now explain Step 2 of the encoding of $t$
which is to obtain from the grammar ${\mathbb{G}}_t$ a string $B({\mathbb{G}}_t) \in {\cal B}$ which is taken as the codeword
$\phi_e(t)$ of $t$.  We will be employing two sequences $S(t)$ and $S_1(t)$ defined as follows:
\begin{itemize}
\item Let $N=N(t)$. For each $i=0,\cdots,N-2$, let ordered pair $(a_{2i+1},a_{2i+2})$ be the right member
of the production rule of ${\mathbb{G}}_t$ whose left member is $i$.
Then $S(t)$ is the sequence
of length $2N-2$ defined by
$$S(t) \define (a_1,a_2,\cdots,a_{2N-3},a_{2N-2}).$$
The alphabet of $S(t)$ is $A(t)=\{1,2,\cdots,N-2\}\cup \{T\}$. Note that ${\mathbb{G}}_t$ is fully
recoverable from $S(t)$.
\item $S_1(t)$ is the sequence of length  $N$ remaining after one deletes from $S(t)$
the first left-to-right appearance in $S(t)$ of each member of the set $\{1,2,\cdots,N-2\}$.
\end{itemize}
Note that $N=N(t)=2$ if and only if $t$ is the unique tree in ${\cal T}_2$; in this case,
${\mathbb{G}}_t$ has only one production rule $0 \to ({ T},{ T})$,
and $S(t)=S_1(t)=({ T},{ T})$.
If $N=2$, define $B({\mathbb{G}}_t)=1$. Now assume
 $N>2$.
The codeword $B({\mathbb{G}}_t)$ will be obtained via processing of the sequence $S(t)$.
Note that $S(t)$ partitions into the two subsequences $S_1(t)$ (defined previously)
and $S_2(t)=(1,2,\cdots,N-2)$.
For each $a \in A(t)$, define $f_a$ to be the positive integer
$$f_a \define {\rm card}\{1 \leq i \leq 2N-2: a_i=a\},$$
that is, $(f_a: a\in{A}(t))$ is the un-normalized first-order empirical distribution
of $S(t)$.
Let ${\cal S}_1(t)$ be the set of all possible permutations of $S_1(t)$; the cardinality
of ${\cal S}_1(t)$ is then computable as
$${\rm card}({\cal S}_1(t)) = \frac{N!}{f_T!\prod_{i=1}^{N-2}(f_i-1)!}.$$
$B({\mathbb{G}}_t)$ is defined to be the left-to-right concatenation of
the binary strings $B_1,B_2,B_3,B_4$ obtained as follows:
\begin{itemize}
\item $B_1$ is the binary string of length $N-1$ consisting of $N-2$ zeroes followed by $1$.
\item $B_2$ is the binary string of length $2N-2$ in which there are exactly $N-2$ entries
equal to $1$, where these entries correspond to the first left-to-right appearances in $S(t)$ of the members
of the set $\{1,2,\cdots,N-2\}$. Given $B_2$, one can reconstruct $S(t)$ from its two
subsequences $S_1(t)$ and $S_2(t)$.
\item $B_3$ is the binary string consisting of $N-1$ alternate runs of ones and zeroes, where the lengths of the
runs (left-to-right) are taken to be $f_1,f_2,\cdots,f_{N-2},1$, respectively. Since $f_T>1$, $B_3$
is of length less than $2N-2$.
\item Let $M(t)=\lceil\log_2{\rm card}({\cal S}_1(t))\rceil$. If $M(t)=0$, $B_4$ is the empty string.
Otherwise, list all members of ${\cal S}_1(t)$ in the lexicographical ordering resulting from the ordering
$1,\cdots,N-2,T$ of the alphabet $A(t)$. Assign each member of the list an index, starting with index $0$. Let $I$ be the index of $S_1(t)$
in this list. $B_4$ is the length $M(t)$ binary expansion of integer $I$.
\end{itemize}
\par

{\it Verification of Prefix Condition.} Suppose $t\in {\cal T}$ has been processed by the encoding map $\phi_e$ to
yield codeword $\phi_e(t)=B({\mathbb{G}}_t)$. Step 1 of the decoding map $\phi_d$ is to determine the
grammar ${\mathbb{G}}_t$ from $B({\mathbb{G}}_t)$. More generally,
we discuss here how $S(t)$ and hence ${\mathbb{G}}_t$  is recoverable from any binary word $w$
of which codeword $B({\mathbb{G}}_t)=B_1B_2B_3B_4$ is a prefix; this will establish that the encoding map $\phi_e:{\cal T}\to{\cal B}$
satisfies the prefix condition.  Scanning $w$
left-to-right to find the first $1$, one determines $B_1$ and $N=N(t)$. $B_2$ is then determined from the fact that its
length is $2N-2$, and then $B_3$ is determined from the fact that it consists of $N-1$ runs.
Knowledge of $B_3$ allows one to determine the set ${\cal S}_1(t)$ and to compute $M(t)$, the length of $B_4$,
whence $B_4$ can be extracted from $w$. From $B_4$, one is able to locate
$S_1(t)$ in the list of the members of ${\cal S}_1(t)$. Using $B_2$, one is able
to put together $S(t)$ from $S_1(t)$ and $S_2(t)$.

\subsection{Encoding/Decoding Examples}

We present two examples. Example 4 illustrates how the encoding map $\phi_e$ works, and Example 5
illustrates how the decoding map $\phi_d$ works. \par

{\it Example 4.} Let $t$ be the tree on the right in Fig.\ 1. Fig.\ 3 illustrates the results
of Step 1 of encoding map $\phi_e$.
We then obtain
$$N=N(t)=8,$$
$$S(t) = (1,2,3,4,3,T,5,4,T,6,6,T,T,T),$$
$$S_1(t) = (3,T,4,T,6,T,T,T),$$
$$S_2(t) = (1,2,3,4,5,6),$$
$$f_1=f_2=f_5=1,\;f_3=f_4=f_6=2,\;f_T=5,$$
$$B_1=0000001,$$
$$B_2=11110010010000,$$
$$B_3=1011001001.$$
We now list the $8!/5!=336$ members of ${\cal S}_1(t)$ in lexicographical order until  $S_1(t)$ is obtained:

$$\begin{array}{|c|c||c|c|}
\hline{\rm index}&{\rm sequence}&{\rm index}&{\rm sequence}\\
\hline 0& (3,4,6,T,T,T,T,T) & 7& (3,6,T,4,T,T,T,T)\\
\hline 1& (3,4,T,6,T,T,T,T) &  8& (3,6,T,T,4,T,T,T)\\
\hline 2& (3,4,T,T,6,T,T,T) &  9& (3,6,T,T,T,4,T,T)\\
\hline 3& (3,4,T,T,T,6,T,T) & 10& (3,6,T,T,T,T,4,T)\\
\hline 4& (3,4,T,T,T,T,6.T) & 11& (3,6,T,T,T,T,T,4)\\
\hline 5& (3,4,T,T,T,T,T,6) & 12& (3,T,4,6,T,T,T,T)\\
\hline 6& (3,6,4,T,T,T,T,T) & 13& (3,T,4,T,6,T,T,T)\\
\hline
\end{array}$$
The index of $S_1(t)$ is thus $I=13$. (Alternatively,
one can use the method of Cover \cite{cover}
to compute $I$ directly without forming the above list.)
To obtain $B_4$, we expand the index $I=13$ into its $\lceil\log_2336\rceil=9$ bit
binary expansion, which yields
$$B_4 = 000001101.$$
The codeword $\phi_e(t) = B_1B_2B_3B_4$ is of length $7+14+10+9 = 40$.
\par

{\it Example 5.} Let binary tree  $t\in {\cal T}$ be such that
$$\phi_e(t) = B({\mathbb{G}}_t) = 00011101000010011000001.$$
We employ the decoding map $\phi_d$ to find $t$ from $B({\mathbb{G}}_t)$. In Decoding Step 1, the grammar ${\mathbb{G}}_t$ must be determined, which,
as remarked earlier, is equivalent to finding the sequence $S(t)$. $B({\mathbb{G}}_t)=B_1B_2B_3B_4$
must be parsed its constituent parts $B_1, B_2, B_3, B_4$.
$B_1$ is the unique prefix of $B({\mathbb{G}}_t)$ belonging to the set $\{1,01,001,001,0001,\cdots\}$,
whence $B_1 = 0001,$
and hence $N=N(t)=4+1=5$.
Thus, $S(t)$ and $B_2$ are both of length $2N-2=8$, whence
$$B_2=11010000$$
and $S(t)$ is of the form
$$S(t)=(a_1,a_2,a_3,a_4,a_5,a_6,a_7,a_8).$$
The positions of symbol $1$ in $B_2$ tell us that
$$S_2(t) = (a_1,a_2,a_4) = (1,2,3),$$
and therefore $S_1(t)$ is made up of the remaining entries in $S(t)$, giving us
$$S_1(t) = (a_3,a_5,a_6,a_7,a_8).$$
Since $B_3$ consists of $N-1=4$ runs of ones and zeroes, with the last run of length $1$, we must have
$$B_3 = 100110.$$
The alphabet of $S(t)$ is $\{1,2,\cdots,N-2,{ T}\}=\{1,2,3,{ T}\}$, and so from $B_3$ the
frequencies of $1,2,3$ in $S(t)$ are the lengths of the first three runs in $B_3$, respectively,
whence
$$f_1 = 1,\;\;f_2 = 2,\;\; f_3 = 2.$$
The remaining entries of $S(t)$ are all equal to ${T}$, giving us $f_{ T}=8-(1+2+2)=3$. It follows
that $S_1(t)$ consists of $f_1-1=0$ entries equal to $1$, $f_2-1=1$ entry equal to $2$,
$f_3-1=1$ entry equal to $3$, and $f_{ T}=3$ entries equal to $T$. Consequently,
${\cal S}_1(t)$ is the set of all permutations of $(2,3,{ T},{ T},{ T})$. The cardinality
of this set is $5!/3! = 20$, and so $B_4$ is of length $\lceil\log_220\rceil=5$. This checks
with what is left of $B({\mathbb{G}}_t)=B_1B_2B_3B_4$ after $B_1,B_2,B_3$ are removed, namely
$$B_4 = 00001.$$
The index of $S_1(t)$ in the list of the members of ${\cal S}_1(t)$ is thus $I=1$. This list starts with
$(2,3,{ T},{ T},{ T})$, which has index $0$, and the sequence following this must therefore by $S_1(t)$.
We conclude that
$$S_1(t) = (2,{ T},3,{ T},{ T}).$$
$S_1(t)$ and $S_2(t)$ now both being known, we put them together to obtain
$$S(t) = (1,2,2,3,{ T},3,{ T},{ T}).$$
Partitioning $S(t)$ into blocks of length two, we obtain  the
four production rules of ${\mathbb{G}}_t$ in Fig.\ 3, whereupon ${\mathbb{G}}_t$ is determined, completing Decoding Step 1. In Decoding Step 2,
one grows the derivation tree of ${\mathbb{G}}_t$ from the production rules of ${\mathbb{G}}_t$  as explained in the Introduction,
giving us the derivation tree in Fig.\ 3; stripping the labels from this tree, we obtain
the binary tree $t$ on the left in Fig.\ 1, completing Decoding Step 2.

\subsection{Performance Bound}
We present Theorem 1, which gives us an upper bound on the
lengths of the binary codewords assigned by the encoding map $\phi_e$ which shall be useful
in later sections. Theorem 1 uses the notion of the first order empirical probability
distribution of a sequence $(s_1,s_2,\cdots,s_n)$ whose entries are selected
from a finite alphabet $A$, which is the probability distribution $p=(p_a:a\in A)$
defined by
$$p_a \define n^{-1}{\rm card}\{1 \leq i \leq n: s_i=a\},\;\;a\in A.$$
The Shannon entropy $H(p)$ of this first order empirical distribution $p$ is defined as
$$H(p) \define \sum_{a\in A}-p_a\log_2p_a,$$
which is also expressible as
$$H(p) = n^{-1}\sum_{i=1}^n-\log_2p_{s_i}.$$
\par

{\bf Theorem 1.} Let $t$ be any binary tree in $\cal T$. Let $p_t$ be the first
order empirical probability distribution of the sequence ${S}_1(t)$. Then
\begin{equation}
L[\phi_e(t)] \leq 5(N(t)-1) +  N(t)H(p_t).
\label{mar19eq1}\end{equation}
\par
{\it Proof.} Let $N=N(t)$. We have $N\geq 2$. If $N=2$, then
$t$ is the unique tree in ${\cal T}_2$ and $L[\phi_e(t)]=1$, whence (\ref{mar19eq1}) holds because the
right side is $5$.
Assume $N>2$. Recall that ${\cal S}_1(t)$ is the set of all permutations of $S_1(t)$.
From the relationships
$$L[\phi_e(t)] = \sum_{i=1}^4L[B_i] = 3(N-1) + L[B_3] + \lceil\log_2({\rm card}({\cal S}_1(t)))\rceil,$$
$$L[B_3] \leq 2N-3,$$
$$\lceil\log_2({\rm card}({\cal S}_1(t)))\rceil \leq \log_2({\rm card}({\cal S}_1(t))) + 1,$$
we obtain
$$L[\phi_e(t)] \leq 5(N-1) + \log_2({\rm card}({\cal S}_1(t))).$$
Since ${\cal S}_1(t)$ is a type class of sequences of length $N$ in the sense of Chapter 2 of \cite{imre}, Lemma 2.3 of \cite{imre}
tells us that
$$\log_2({\rm card}({\cal S}_1(t))) \leq NH(p_t).$$
Inequality (\ref{mar19eq1}) is now evident.

\subsection{Minimality/Uniqueness of ${\mathbb{G}}_t$}

Given $t\in {\cal T}$, we discuss what distinguishes ${\mathbb{G}}_t$ among the possibly many deterministic context-free grammars which
form a representation of $t$. First, we explain what it means for a directed acyclic graph (DAG) to be a representation of $t$.
Let $D$ be a finite rooted DAG with at least two vertices such that each non-leaf vertex has exactly two ordered edges. Define ${\mathbb{G}}(D)$ to be the deterministic context-free grammar whose set of nonterminal variables is the set of non-leaf vertices of $D$, whose set of terminal variables is the set of leaf vertices of $D$, whose
start variable is the root vertex of $D$, and whose production rules are all the rules of the form $v\to (v_1,v_2)$ in which $v$
is a non-leaf vertex of $D$, and $v_1,v_2$ are the respective vertices of $D$ at the terminus of the edges $1,2$ emanating from $v$.
Then we say that $D$ is a representation of $t\in\cal T$ if the grammar ${\mathbb{G}}(D)$ forms a representation of $t$.
It is known that each binary tree in $\cal T$ has a unique DAG representation up to isomorphism
with the minimal number of vertices \cite{maneth1}; we call this DAG the minimal DAG representation
of the binary tree. One particular choice
of minimal DAG representation of $t \in {\cal T}$ is the DAG $D^*(t)$ defined as follows. The set of vertices of
$D^*(t)$ is $\{t(v):v\in V(t)\}$. The root vertex of $D^*(t)$ is $t$, and $t^*$ is the unique
leaf vertex of $D^*(t)$. If $u$ is a non-leaf vertex of $D^*(t)$, then there are exactly two ordered edges
emanating from $u$, edge $1$ terminating at $u_L$ and edge $2$ terminating at $u_R$. Note that
the number of vertices of the minimal DAG representation $D^*(t)$ of $t$ is $N(t)$, which coincides
with the number of variables of ${\mathbb{G}}_t$. (Recall that the complete set of variables of ${\mathbb{G}}_t$
is $\{0,1,\cdots,N(t)-2\}\cup\{T\}$, of cardinality $N(t)$.) The paper \cite{maneth2} gives a linear-time algorithm
for computing $D^*(t)$. Fig.\ 4 illustrates a binary tree together with its minimal DAG representation.
\par

{\bf Lemma 1.} Let $t \in {\cal T}$. Then ${\mathbb{G}}_t$ has the smallest number of variables among all deterministic context-free grammars
which form a representation of $t$.
\par
{\it Proof.} Let ${\mathbb{G}}$ be a deterministic context-free grammar which forms a representation of $t$.
The proof consists in showing that the number of variables of ${\mathbb{G}}$ is at least $N(t)$, the number of variables of ${\mathbb{G}}_t$.
In the following, we explain how to extract from the derivation tree $t({\mathbb{G}})$ of ${\mathbb{G}}$
a rooted ordered DAG $D(t)$ which is a representation of $t$. The set of vertices of $D(t)$ is the
set of labels on the vertices of $t({\mathbb{G}})$. The root vertex of $D(t)$ is the label on the root vertex of $t({\mathbb{G}})$,
the set of non-leaf vertices of $D(t)$ is the set of labels on the non-leaf vertices of $t({\mathbb{G}})$, and
the set of leaf vertices of $D(t)$ is the set of labels on the leaf vertices of $t({\mathbb{G}})$. Let $s$
be any non-leaf vertex of $D(t)$. Find a vertex $v$ of $t({\mathbb{G}})$ whose label is $s$, and let $s_1,s_2$
be the respective labels on the ordered children of $v$ in $t({\mathbb{G}})$; the pair $(s_1,s_2)$ thus derived
will be the same no matter which vertex $v$ of $t({\mathbb{G}})$ with label $s$ is chosen. There are exactly
two ordered edges of $D(t)$ emanating from $s$, namely, edge $1$ which terminates at $s_1$ and edge $2$ which terminates
at $s_2$. This completes the specification of the DAG $D(t)$. By construction of $D(t)$, the number of variables of ${\mathbb{G}}$
is at least as much as the number of vertices of $D(t)$. Since $D(t)$ is a DAG representation of $t$, the number
of vertices of $D(t)$ is at least as much as the number of vertices $N(t)$ of the minimal DAG
representation of $t$. Thus, the number of variables of ${\mathbb{G}}$ is at least $N(t)$, completing the proof.
\par
{\bf Remark.} With some more work, one can show that  any deterministic context-free grammar which forms a representation of $t\in{\cal T}$ and has the same number of variables
as ${\mathbb{G}}_t$ must be isomorphic to  ${\mathbb{G}}_t$, using the known fact mentioned earlier that the minimal DAG representation of $t$ is unique up to isomorphism.
This gives us a sense in which ${\mathbb{G}}_t$ is unique.

\par

\begin{figure}[htp]
\centering
\fbox{\includegraphics[scale=0.6]{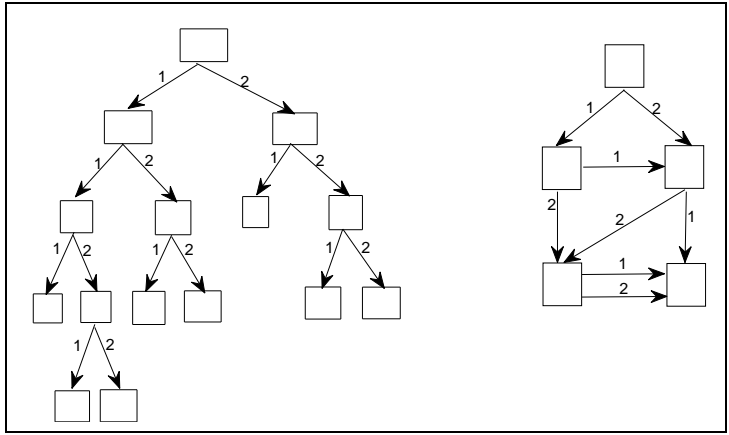}}
\caption{A binary tree (left) and its minimal DAG representation (right)}
\end{figure}
\par

\section{Sources For Which $(\phi_e,\phi_d)$ is Asymptotically Optimal}

This section examines the asymptotic performance of the code $(\phi_e,\phi_d)$ on a binary tree source.
We put forth weak sufficient conditions on a binary tree source
so that our two-step grammar-based code $(\phi_e,\phi_d)$ will be an asymptotically optimal code for the source.
Before doing that, we need to first establish a lemma giving an asymptotic average redundancy
lower bound  for general structure sources.

\par

Suppose $(\Omega,{\cal F},P)$ be an arbitrary structure source. Let $(\psi_e,\psi_d)$ be a lossless code on $\Omega$,
and let $F\in {\cal F}$ be such that every structure $\omega \in F$ is of the same size.
The well-known entropy lower bound for prefix codes tells us that
$$\sum_{\omega\in F}L[\psi_e(\omega)]P(\omega) \geq \sum_{\omega\in F,\;P(\omega)>0}-P(\omega)\log_2P(\omega),$$
from which it follows that
$$R(\psi_e,F,P) \geq 0,$$
that is, the $F$-th order average redundancy of the code with respect to the source is non-negative.
Although this redundancy non-negativity property fails
for a general structure source, the following result gives us an asymptotic sense
in which average redundancy is non-negative.
\par

{\bf Lemma 2.} Let $(\Omega,{\cal F},P)$ be a general structure source. Then
\begin{equation}
\liminf_{F\in{\cal F}}R(\psi_e,F,P) \geq 0
\label{apr16eq2}\end{equation}
for any lossless code $(\psi_e,\psi_d)$ on $\Omega$.

\par
{\it Proof.} Fix a general structure source $(\Omega,{\cal F},P)$. Let $\mathbb{Q}$ be the set of all $Q:\Omega\to(0,1)$
such that the restriction of $Q$ to each $F\in{\cal F}$ is a probability distribution on $F$.
In the first part of the proof, we show that
\begin{equation}
\liminf_{F\in{\cal F}}\sum_{\omega\in{ F}}|\omega|^{-1}P(\omega)\log_2\left(\frac{P(\omega)}{Q(\omega)}\right) \geq 0,\;\;Q\in{\mathbb{Q}},
\label{apr16eq3}\end{equation}
where in (\ref{apr16eq3}) and henceforth, any expected value of the form $\sum_{\omega\in F}g(\omega)P(\omega)$
is computed by summing only over those  $\omega\in F$ for which $P(\omega)>0$.
The proof of (\ref{apr16eq3}) exploits the concept of divergence. If $p=(p_j:j \in A)$ and $q=(q_j:j\in A)$ are any two probability distributions on a finite set $A$, with all $q_j$ probabilities $>0$,
we let $D(p|q)$ denote the
divergence of $p$ with respect to $q$, defined by
$$D(p|q) \define \sum_{j\in A}p_j\log_2\left(\frac{p_j}{q_j}\right).$$
It is well-known that $D(p|q)\geq 0$ \cite{imre}. Fix an arbitrary $Q \in {\mathbb{Q}}$.
Given $F \in {\cal F}$, let $I_F=\{|\omega|: \omega\in F\}$, and for each $i\in I_F$, let $F_i=\{\omega\in F:|\omega|=i\}$.
Furthermore,
 let $P_F,Q_F$ be the probability distributions on $I_F$ such that
$$P_F(i) = P(F_i),\;\;i\in I_F,$$
$$Q_F(i) = Q(F_i),\;\;i\in I_F,$$
and for each $i \in I_F$, let $P_F^i,Q_F^i$ be probability distributions on $F_i$ such that
$$P(\omega)=P_F(i)P_F^i(\omega),\;\;\omega \in F_i,$$
$$Q(\omega)=Q_F(i)Q_F^i(\omega),\;\;\omega\in F_i.$$
It is easy to show that
$$\sum_{\omega\in{ F}}|\omega|^{-1}P(\omega)\log_2\left(\frac{P(\omega)}{Q(\omega)}\right) = \sum_{i\in I_F}i^{-1}P_F(i)D(P_F^i|Q_F^i) + \sum_{i\in I_F}i^{-1}P_F(i)\log_2\left(\frac{P_F(i)}{Q_F(i)}\right),$$
and therefore
$$\sum_{\omega\in{ F}}|\omega|^{-1}P(\omega)\log_2\left(\frac{P(\omega)}{Q(\omega)}\right) \geq \sum_{i\in I_F}i^{-1}P_F(i)\log_2\left(\frac{P_F(i)}{Q_F(i)}\right).$$
Let $E_P^F, E_Q^F$ be the expected values defined by
$$E_P^F \define \sum_{i\in I_F}i^{-1}P_F(i),$$
$$E_Q^F \define \sum_{i\in I_F}i^{-1}Q_F(i).$$
Note that $E_P^F$ and $E_Q^F$ both belong to the interval $(0,1]$.
Let $P_F^*,Q_F^*$ be the probability distributions on $I_F$ defined by
$$P_F^*(i) \define i^{-1}P_F(i)/E_P^F,\;\;i\in I_F,$$
$$Q_F^*(i) \define i^{-1}Q_F(i)/E_Q^F,\;\;i\in I_F.$$
Then we have
$$\sum_{i\in I_F}i^{-1}P_F(i)\log_2\left(\frac{P_F(i)}{Q_F(i)}\right) = E_P^FD(P_F^*|Q_F^*) + E_P^F\log_2(1/E_Q^F) + E_P^F\log_2E_P^F.$$
Since $1/E_Q^F \geq 1$, the first two terms on the right side of the preceding equality are non-negative, whence
\begin{equation}
\liminf_{F\in{\cal F}}\sum_{\omega\in{F}}|\omega|^{-1}P(\omega)\log_2\left(\frac{P(\omega)}{Q(\omega)}\right) \geq \liminf_{F\in{\cal F}}E_P^F\log_2E_P^F.
\label{apr16eq5}\end{equation}
Note that
$$0 < E_P^F \leq \frac{1}{\min\{|\omega|:\omega\in F\}},$$
and so by (\ref{mar28eq2})
\begin{equation}
\lim_{F\in{\cal F}}E_P^F = 0,
\label{apr16eq6}\end{equation}
the right side of (\ref{apr16eq5}) is zero, and (\ref{apr16eq3}) holds.
To finish the proof, let $(\psi_e,\psi_d)$ be any lossless code on $\Omega$. By Kraft's inequality for prefix codes,
there exists $Q \in {\mathbb{Q}}$ such that
$$L[\psi_e(\omega)] \geq -\log_2Q(\omega),\;\;\omega\in\Omega,$$
and hence
$$R(\psi_e,F,P) = \sum_{\omega\in F}|\omega|^{-1}\{L[\phi_e(\omega)]+\log_2P(\omega)\}P(\omega) \geq \sum_{\omega\in{ F}}|\omega|^{-1}P(\omega)\log_2\left(\frac{P(\omega)}{Q(\omega)}\right).$$
(\ref{apr16eq2}) then follows by appealing to (\ref{apr16eq3}).
\par

{\bf Remark.} In view of Lemma 2, given a general structure source $(\Omega,{\cal F},P)$, a lossless code $(\psi_e,\psi_d)$
on $\Omega$ is an asymptotically optimal code for the source if and only if
\begin{equation}
\limsup_{F\in {\cal F}}R(\psi_e,F,P) \leq 0.
\label{apr23eq1}\end{equation}
\par

We now turn our attention to properties of a binary tree source under which the grammar-based code $(\phi_e,\phi_d)$ on $\cal T$ will be
asymptotically optimal for the source. There are two of these properties, the Domination Property and the Representation Ratio
Negligibility Property, which are discussed in the following.\par

{\it Domination Property.} We define $\Lambda$ to be the set of all mappings $\lambda:{\cal T}^*\to (0,1]$ such that
\begin{itemize}
\item {\bf (a):} $\lambda(t) \leq \lambda(t_L)\lambda(t_R),\;\;t\in {\cal T}$.
\item {\bf (b):} There exists a positive integer $K(\lambda)$ such that
\begin{equation}
1 \leq \sum_{t\in{\cal T}_n}\lambda(t)\leq n^{K(\lambda)},\;\;n\geq 1.
\label{mar21eq2}\end{equation}
\end{itemize}
An element $\lambda$ of $\Lambda$ dominates a binary tree source $({\cal T},{\cal F},P)$  if
$P(t) \leq \lambda(t)$ for all $t\in{\cal T}$. A binary tree source satisfies the Domination Property
if there exists an element of $\Lambda$ which dominates the source.
\par

{\it Representation Ratio Negligibility Property.} Let $t\in{\cal T}$. We define the representation ratio of $t$,
denoted $r(t)$, to be the
ratio between the number of variables of the grammar ${\mathbb{G}}_t$
and the number of  leaves of $t$. That  is, $r(t)=N(t)/|t|$.
Since
$$N(t) = {\rm card}\{t(v):v\in V(t)\}=1+{\rm card}\{t(v):v\in V^1(t)\} \leq 1+(|t|-1)=|t|,$$
the representation ratio is at most $1$.
In the main result of this section, Theorem 2, we will see that our ability
to compress $t \in {\cal T}$ via the code $(\phi_e,\phi_d)$ becomes greater as
$r(t)$ becomes smaller. We say that a binary tree source $({\cal T},{\cal F}, P)$ obeys the Representation
Ratio Negligibility Property (RRN Property) if
\begin{equation}
\lim_{F\in{\cal F}}\sum_{t\in F}r(t)P(t) = 0.
\label{apr5eq5}\end{equation}
\par

{\it Definition.} Henceforth, $\gamma:[0,1]\to [0,\infty)$ is the function defined by
\[ \gamma(x) \define \left \{ \begin{array} {r@{\quad}l}
-(x/2)\log_2(x/2), & x>0\\
0, & x=0
\end{array} \right. \]
\par

{\bf Theorem 2.} The following statements hold:
\begin{description}
\item{{\bf (a):}} For each $\lambda\in\Lambda$,
\begin{equation}
|t|^{-1}\left\{L[\phi_e(t)]+\log_2\lambda(t)\right\} \leq (2K(\lambda)+10)\gamma(r(t)),\;\;t\in {\cal T}.
\label{mar21eq1}\end{equation}
\item{{\bf (b):}} Let $({\cal T},{\cal F},P)$ be a
binary tree source satisfying the Domination Property, where $\cal F$ can be any ${\cal T}$-filter. There exists a positive real number $C$, depending only on
the source, such that
\begin{equation}
R(\phi_e,F,P) \leq C\gamma\left(\sum_{t\in{F}}r(t)P(t)\right),\;\;F\in{\cal F}.
\label{apr17eq1}\end{equation}
\item{{\bf (c):}} $(\phi_e,\phi_d)$ is an asymptotically optimal code for any binary tree source which satisfies both the
Domination Property and the RRN Property.
\end{description}
\par

{\it Proof.} It suffices to prove part (a). (Part (b) follows from part (a) and the
fact that $\gamma$ is a concave function; part(c) follows from part(b) and (\ref{apr23eq1}).) Let $\lambda \in \Lambda$ be arbitrary.
Fix $t\in{\cal T}$ and let $N=N(t)$.
 There is an initial binary subtree $t^{\dagger}$ of $t$ such that
\begin{itemize}
\item There are $N$ leaf vertices of $t^{\dagger}$.
\item The subtrees $t(v)$ are distinct as $v$ ranges through the $N-1$ non-leaf vertices of $t^{\dagger}$.
\end{itemize}
(One can obtain $t^{\dagger}$ either by pruning the derivation tree of ${\mathbb{G}}_t$ or by growing it using
the production rules of ${\mathbb{G}}_t$ so that in the growth process each production rule is used to extend a leaf exactly
once; see Fig.\ 5.)
Let $v_1,v_2,\cdots,v_N$ be an enumeration of
the leaves of $t^{\dagger}$. There is a
one-to-one correspondence between the set $\{t(v):v\in V(t)\}$ and the
set of variables of ${\mathbb{G}}_t$, and under this correspondence, the
sequence $s^*=(t(v_1),t(v_2),\cdots,t(v_N))$ is carried into a sequence which is
a permutation of the sequence $S_1(t)$, and the first order empirical distribution
$p^*$ of $s^*$ is carried into the first order empirical distribution $p_t$ of $S_1(t)$.
Thus, the Shannon entropies $H(p^*)$, $H(p_t)$ coincide, and appealing to Theorem 1, we have
$$L[\phi_e(t)] \leq 5(N-1) + \sum_{i=1}^N-\log_2p^*(t(v_i)).$$
Define
$$M_j \define \sum_{u\in{\cal T}_j}\lambda(u),\;\;j\geq 1.$$
There is a unique real number $D > 1/2$ such that
\begin{equation}
q(u) \define DM_j^{-1}|u|^{-2}\lambda(u),\;\;u\in{\cal T}_j,\;\;j\geq 1
\label{mar4eq7}\end{equation}
defines a probability distribution on ${\cal T}^*$. Shannon's Inequality (\cite{aczel}, page 37) then gives us
$$\sum_{i=1}^N-\log_2p^*(t(v_i)) \leq \sum_{i=1}^N-\log_2q(t(v_i)).$$
Using formula (\ref{mar4eq7}) and the fact that $-\log_2D\leq 1$, we obtain
\begin{eqnarray*}
\sum_{i=1}^N-\log_2q(t(v_i)) &=& N(-\log_2D) + Q_1 + 2Q_2 + Q_3\\
&\leq & N + Q_1+2Q_2+Q_3,
\end{eqnarray*}
where
\begin{eqnarray*}
Q_1 &=& \sum_{i=1}^N\log_2M_{|t(v_i)|},\\
Q_2 &=& \sum_{i=1}^N \log_2|t(v_i)|,\\
Q_3 &=& -\sum_{i=1}^N\log_2\lambda(t(v_i)).
\end{eqnarray*}
We bound each of these quantities in turn. By (\ref{mar21eq2}), we obtain
$$Q_1 \leq K(\lambda)Q_2.$$
By concavity of the logarithm function, and recalling that $r(t)=N/|t|$, we have
$$Q_2 \leq N\log_2\left(\frac{\sum_{i=1}^N|t(v_i)|}{N}\right) = N\log_2(|t|/N) = 2|t|\gamma(r(t))-N.$$
By property (a) for membership of $\lambda$ in $\Lambda$, we have
$$Q_3 \leq -\log_2\lambda(t).$$
Combining previous bounds, and writing $K=K(\lambda)$, we see that
\begin{eqnarray*}
L[\phi_e(t)] + \log_2\lambda(t) &\leq& 6N-(K+2)N +2(K+2)|t|\gamma(r(t))\\
                                 &\leq &3|t|r(t)+2(K+2)|t|\gamma(r(t))
                                 \end{eqnarray*}
holds, whence (\ref{mar21eq1}) holds because $r(t) \leq 2\gamma(r(t))$, completing the proof
of part (a) of Theorem 2.
\par
\begin{figure}[htp]
\centering
\fbox{\includegraphics[scale=0.5]{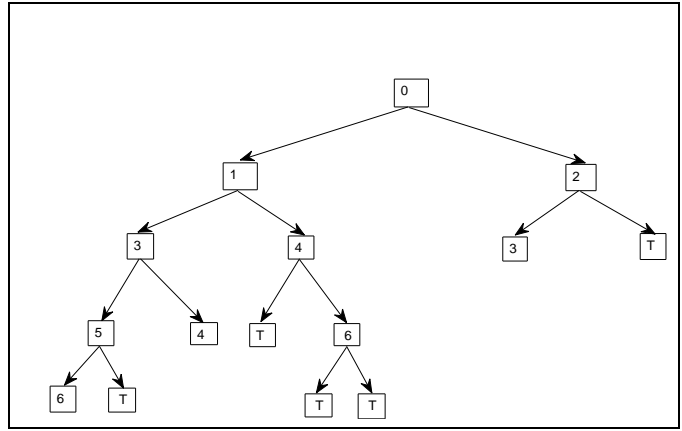}}
\caption{Initial subtree of Fig.\ 3 derivation tree used in Theorem 2 proof}
\end{figure}

\section{Universal Coding of Leaf-Centric Binary Tree Sources}

We fix throughout this section the ${\cal T}$-filter ${\cal F}_1=\{{\cal T}_n:n\geq 2\}$. We now formally
define the set of leaf-centric binary tree sources, which are certain binary tree sources of the form
$({\cal T},{\cal F}_1,P)$. Let $\mathbb{N}$ be the set of positive integers, and let $\Sigma_1$ be the set of all functions
 $\sigma$ from $\mathbb{N}\times \mathbb{N}$
into $[0,1]$ such that
$$\sum_{\{(i,j):i,j\geq 1,\;i+j=n\}} \sigma(i,j)=1,\;\;n\geq 2.$$
For each $\sigma \in {\Sigma}_1$, let $P_{\sigma}$ be the mapping from ${\cal T}$ into $[0,1]$
  such that
  $$P_{\sigma}(t) = \prod_{v\in V^1(t)}\sigma(|t(v)_L|,|t(v)_R|),\;\;t\in {\cal T}.$$
Since
$$\sum_{t\in{\cal T}_n}P_{\sigma}(t) = 1,\;\;n\geq 2,$$
$S(\sigma)=({\cal T},{\cal F}_1,P_{\sigma})$ is a binary tree source.
The sources in the family $\{S(\sigma):\sigma\in\Sigma_1\}$ are called leaf-centric binary tree sources,
the reason being that  the probability of each  tree
is computed based purely upon the number of leaves in each of its final subtrees. Leaf-centric
binary tree sources were first considered in the paper \cite{isit2009}.
\par

{\it Example 6.} Let $\Sigma_1^{\dagger}$ be the subset of $\Sigma_1$ consisting of all $\sigma\in\Sigma_1$
for which
$$\{(i,j):i,j\geq 1,\;i+j=n,\;\sigma(i,j)>0\} \subset \{(1,n-1),(n-1,1)\},\;\;n\geq 2.$$
If $\sigma\in\Sigma_1^{\dagger}$, then a tree $t\in{\cal T}$ with positive $P_{\sigma}$ probability
must satisfy the property that there exist only two vertices of $t$ at each depth level of $t$ beyond
level $0$; we call such a binary tree a one-dimensional tree. Consider the structure
universe of binary strings $\cal B$, in which the size of a string $b\in{\cal B}$ is taken to be its length $L[b]$.
For each $n\geq 1$, let ${\cal B}_n$ be the set of strings in $\cal B$ of length $n$, and let
${\cal F}({\cal B})$ be the ${\cal B}$-filter $\{{\cal B}_n:n\geq 1\}$.
Let $[0,1]^{\infty}$ be the set of all sequences
$\alpha=(\alpha_i:i\geq 1)$ in which each $\alpha_i$ belongs to the interval $[0,1]$,
and for each $\alpha\in [0,1]^{\infty}$,
let $({\cal B},{\cal F}({\cal B}),Q_{\alpha})$ be the one-dimensional source
in which for each string $b_1b_2\cdots b_n$ belonging to ${\cal B}$ we have
$$Q_{\alpha}(b_1b_2\cdots b_n) = \prod_{i=1}^nq(\alpha_i,b_i),$$
where $q(\alpha_i,b_i)$ is taken to $\alpha_i$ if $b_i=0$ and taken to be $1-\alpha_i$, otherwise.
It is easy to see that the family of sources $\{({\cal T},{\cal F}_1,P_{\sigma}):\sigma\in\Sigma_1^{\dagger}\}$ has
a universal code if and only if the family of one-dimensional sources  $\{({\cal B},{\cal F}({\cal B}),Q_{\alpha}):\alpha\in [0,1]^{\infty}\}$
has a universal code. The third author has
shown that this latter family of one-dimensional sources has no universal code. Therefore, the family $\{S(\sigma):\sigma\in\Sigma_1^{\dagger}\}$
has no universal code, and so the bigger family of all leaf-centric binary
tree sources also has no universal code.
\par

The following result shows that $(\phi_e,\phi_d)$ is a universal code for a suitably restricted subfamily of
the family of leaf-centric binary tree sources.
\par

  {\bf Theorem 3.}   Let $\Sigma_1^*$ be the uncountable set consisting of all $\sigma\in\Sigma_1$ such that
  \begin{equation}
  \sup\left\{\frac{i+j}{\min(i,j)}: i,j\geq 1,\;\;\sigma(i,j)>0\right\} < \infty.
  \label{apr4eq5}\end{equation}
 Then $(\phi_e,\phi_d)$ is a universal code for the family of
  sources $\{S(\sigma):\sigma\in\Sigma_1^*\}$.
  \par

  Before proceeding with the proof of Theorem 3, we provide an example of a source in
  $\{S(\sigma):\sigma\in\Sigma_1^*\}$.
  \par

  {\it Example 7.} Given a general structure source $(\Omega,{\cal F},P)$, then for each $F\in{\cal F}$, the
  $F$-th order entropy of the source is defined by
  $$H_F(P) \define \sum_{\omega\in F}-|\omega|^{-1}P(\omega)\log_2P(\omega).$$
  $\lim_{F\in{\cal F}}H_F(P)$ is defined to be the entropy rate of the source, if the limit exists;
  otherwise, the source has no entropy rate.
   In universal source coding theory for families of classical one-dimensional
  sources (see Ex.\ 3), the sources are typically assumed to be stationary sources or finite-state sources, which are types
  of sources which have an entropy rate. In the universal coding of binary tree sources, however, one very often
  deals with sources which have no entropy rate. We illustrate a particular source of this type in the family
  $\{S(\sigma):\sigma\in\Sigma_1^*\}$. Let $\sigma \in {\Sigma}_1^*$ be the function such that for each even $n\geq 2$,
  $$\sigma(n/2,n/2)=1,$$
  and for each odd $n \geq 3$,
  $$\sigma(\lfloor{n/2}\rfloor,\lceil{n/2}\rceil) = \sigma(\lceil{n/2}\rceil,\lfloor{n/2}\rfloor) = 1/2.$$
  The resulting leaf-centric binary tree source $S(\sigma)$,
   introduced in \cite{isit2009}, is called the
  {\it bisection tree source model}.
  In \cite{AofA2012}, it is shown that there is a unique nonconstant continuous periodic function $f:{\mathbb{R}}\to[0,1]$, with
  period $1$, such that
  \begin{equation}
  -\log_2P_{\sigma}(t) = |t|f(\log_2|t|),\;\;t\in{\cal T},
  \label{apr2eq6}\end{equation}
  and the restriction of $f$ to $[0,1] $ is characterized as the attractor of a specific iterated function system on $[0,1]$;
  because of this property, the source $S(\sigma)$ has no entropy rate.
  \par

   {\it Proof of Theorem 3.} If $\sigma\in\Sigma_1$,
   let
   $\lambda:{\cal T}^*\to [0,1]$ be the function
   such that $\lambda(t^*)=1$ and
   $$\lambda(t) = \max(K_n^{-1},P_{\sigma}(t)),\;\;t\in{\cal T}_n,\;\;n\geq 2.$$
   Then $\lambda\in\Lambda$ and $\lambda$ dominates $P_{\sigma}$.
   Thus, every source in the family $\{S(\sigma):\sigma\in\Sigma_1^*\}$ satisfies the Domination Property.
   By Theorem 2,
   our proof will be complete once it is shown that every source in this family
   satisfies the RRN Property. More generally, we show that the RRN Property holds
   for any
   binary tree source $({\cal T},{\cal F},P)$ for which
   \begin{equation}
  \sup_{t\in{\cal T},\;P(t)>0}\left\{\max_{v\in V^1(t)}\left[\frac{|t(v)|}{\min(|t(v)_L|,|t(v)_R|)}\right]\right\} < \infty.
  \label{apr4eq6}\end{equation}
  (The ${\cal T}$-filter ${\cal F}$ in the given source $({\cal T},{\cal F},P)$ need not be equal to ${\cal F}_1$.)
  Let $C$ be a positive integer greater than or equal to the supremum on the left side of (\ref{apr4eq6}).
  Fix $t\in{\cal T}$ for which $P(t)>0$.
    As in the proof of Theorem 2, let $t^{\dagger}$ be an initial binary subtree of $t$ with $N=N(t)$ leaves
 such that $\{t(v):v\in V^1(t^{\dagger})\}=\{t(v):v\in V^1(t)\}$.
Let $v_1,v_2,\cdots,v_N$ be an enumeration of the leaves of $t^{\dagger}$ and for each $i=1,2,\cdots,N$, let
$u_i \in V^1(t^{\dagger})$ be the parent vertex of $v_i$. We have
$$\frac{|t(u_i)|}{|t(v_i)|} \leq C,\;\;i=1,2,\cdots,N,$$
and therefore
$$\frac{|t(u_1)|+|t(u_2)| + \cdots + |t(u_N)|}{|t(v_1)|+|t(v_2)| + \cdots + |t(v_N)|} \leq C.$$
The sum in the denominator is $|t|$, and so
\begin{equation}
\frac{|t(u_1)|+|t(u_2)| + \cdots + |t(u_N)|}{|t|} \leq C.
\label{mar7eq1}\end{equation}
Each $u \in \{u_1,\cdots,u_N\}$ can be the parent of at most two elements of the
set $\{v_1,\cdots,v_N\}$, and so
$${\rm card}(\{u_1,\cdots,u_N\})\geq (1/2){\rm card}(\{v_1,\cdots,v_N\}) = N/2.$$
The mapping $u\to t(u)$ from the set $V^1(t^{\dagger})$ into the set $\{t(v):v\in V^1(t)\}$
is a one-to-one onto mapping (both sets have cardinality $N-1$). Therefore,
\begin{equation}
{\rm card}(\{t(u_1),t(u_2),\cdots,t(u_N)\}) \geq N/2.
\label{mar7eq2}\end{equation}
Let $k=\lceil N/2\rceil$.
We conclude from (\ref{mar7eq1})-(\ref{mar7eq2}) that there
are $k$ distinct trees $t_1,t_2,\cdots,t_k$ in $\cal T$ whose total number of leaves is $\leq |t|C$,
where we suppose that these $k$ trees have been enumerated so that
$$|t_1| \leq |t_2| \leq \cdots \leq |t_k|.$$
Let $t(1),t(2),t(3),\cdots$ be an enumeration of all trees in  $\cal T$ such that
$t(1)$ is the unique tree in ${\cal T}_2$, $t(2),t(3)$ are the two trees in ${\cal T}_3$,
$t(4),t(5),t(6),t(7),t(8)$ are the five trees in ${\cal T}_4$, and so forth. We clearly
have $|t(i)| \leq |t_i|$ for $i=1.\cdots,k$. Therefore,
\begin{equation}
|t(1)| + |t(2)| + \cdots + |t(k)| \leq |t|C.
\label{mar7eq5}\end{equation}
The sequence $m_i=|t(i)|$ can be characterized as the sequence in which
$m_1 = 2$
and for each $j \geq 3$,
$m_i=j$ for all integers $i$ satisfying
$$K_2+K_3+\cdots+K_{j-1} < i \leq K_2+K_3+\cdots+K_j.$$
Define
$$k(M) \define \max\{k \geq 1: m_1+m_2+\cdots+m_k \leq M\},\;\;M \geq 2.$$
Since the sequence $\{K_j:j\geq 2\}$ grows exponentially fast, it follows that
$k(M)/M = O(1/\log_2M)$ by an argument similar to an argument on page 753 of \cite{Kieffer-Yang1}, and hence
\begin{equation}
\lim_{M\to\infty}k(M)/M = 0.
\label{mar7eq6}\end{equation}
From (\ref{mar7eq5}), we have shown that
$$\lceil N(t)/2\rceil \leq k( |t|C)),\;\;t\in{\cal T},\;\;P(t)>0.$$
Dividing both sides by $|t|$ and summing, we then have
\begin{equation}
\sum_{t\in F}r(t)P(t) \leq 2\sum_{t\in F}|t|^{-1}k(|t|C))P(t),\;\;F\in{\cal F}.
\label{apr4eq7}\end{equation}
Let $n_F=\min\{|t|:t\in F\}$, and define
$$\delta(J) \define \sup\{k(j)/j: j\geq J\},\;\;J\geq 2.$$
From (\ref{apr4eq7}), we then have
\begin{equation}
\sum_{t\in{\cal F}}r(t)P(t) \leq 2C\delta(n_FC),\;\;F\in {\cal F}.
\label{apr4eq8}\end{equation}
By (\ref{mar28eq1}), $\lim_{F\in{\cal F}}n_F=\infty$, and we also have $\lim_{J\to\infty}\delta(J)=0$.
Taking the limit along filter $\cal F$ on both sides of (\ref{apr4eq8}), we then obtain (\ref{apr5eq5}),
which is the RRN Property for the source $({\cal T},{\cal F},P)$.

\section{Universal Coding of Depth-Centric Binary Tree Sources}

For each $t\in {\cal T}^*$, define $d(t)$ to be the depth of $t$, which is the number of edges in the longest
root-to-leaf path in $t$. We have $d(t^*)=0$ and as defined in Ex.\ 2, for each $n\geq 1$ we let ${\cal T}^n$
be the set of trees $\{t\in{\cal T}:d(t)=n\}$.
We fix throughout this section the ${\cal T}$-filter ${\cal F}_2=\{{\cal T}^n:n\geq 1\}$. We now formally
define the set of depth-centric binary tree sources, which are certain binary tree sources of the form
$({\cal T},{\cal F}_2,P)$. Let $\mathbb{Z}^+$ be the set of nonnegative integers, and let $\Sigma_2$ be the set of all functions
 $\sigma$ from $\mathbb{Z}^+\times \mathbb{Z}^+$
into $[0,1]$ such that
$$\sum_{\{(i,j):i,j\geq 0,\;\max(i,j)=n-1\}} \sigma(i,j)=1,\;\;n\geq 1.$$
For each $\sigma \in {\Sigma}_2$, let $P_{\sigma}$ be the mapping from ${\cal T}$ into $[0,1]$
  such that
  $$P_{\sigma}(t) = \prod_{v\in V^1(t)}\sigma(d(t(v)_L),d(t(v)_R)),\;\;t\in {\cal T}.$$
Since
$$\sum_{t\in{\cal T}^n}P_{\sigma}(t) = 1,\;\;n\geq 1,$$
$S(\sigma)=({\cal T},{\cal F}_2,P_{\sigma})$ is a binary tree source.
The sources in the family $\{S(\sigma):\sigma\in\Sigma_2\}$ are called depth-centric binary tree sources,
the reason being that  the probability
of each tree is based purely upon the depths of its final subtrees.
\par

{\it Example 8.} Let $\Sigma_2^{\dagger}$ be the subset of $\Sigma_2$ consisting of all $\sigma\in\Sigma_2$
for which
$$\{(i,j):i,j\geq 0,\;\max(i,j)=n-1,\;\sigma(i,j)>0\} \subset \{(0,n-1),(n-1,0)\},\;\;n\geq 1.$$
If $\sigma\in\Sigma_2^{\dagger}$, then a tree $t\in{\cal T}$ has positive $P_{\sigma}$ probability if and only
if $t$ is a one-dimensional tree. The family of sources $\{S(\sigma):\sigma\in\Sigma_2^{\dagger}\}$
has no universal code by the same argument given in Ex.\ 6.
Thus, the bigger family of all depth-centric binary
tree sources also has no universal code.
\par

Our final result shows that $(\phi_e,\phi_d)$ is a universal code for a suitably restricted subfamily of
the family of depth-centric binary tree sources.
\par

  {\bf Theorem 4.}   Let $\Sigma_2^*$ be the uncountable set consisting of all $\sigma\in\Sigma_2$ such that
  \begin{equation}
  \sup\{|i-j|: i,j\geq 0,\;\;\sigma(i,j)>0\} < \infty
  \label{apr3eq5}\end{equation}
  and
  \begin{equation}
  {\rm card}\{|i-j|:i,j\geq 0,\;\max(i,j)=n-1,\;\sigma(i,j)>0\} = 1,\;\;n\geq 1.
  \label{apr4eq1}\end{equation}
  Then $(\phi_e,\phi_d)$ is a universal code for the family of
  sources $\{S(\sigma):\sigma\in\Sigma_2^*\}$.
  \par

  {\it Proof.} Each source in the family $\{S(\sigma):\sigma\in\Sigma_2^*\}$ satisfies the Domination Property, by the same argument given in the
  proof of Theorem 3. Appealing to Theorem 2, our proof will be complete once we verify
  that each source in this  family also satisfies the RRN Property.  Fix the source $S(\sigma)$, where $\sigma\in\Sigma_2^*$.
  By the last part of the proof of Theorem 3, $S(\sigma)$ will satisfy the RRN Property if
  \begin{equation}
  \sup_{t\in{\cal T},\;P_{\sigma}(t)>0}\left\{\max_{v\in V^1(t)}\left[\frac{|t(v)|}{\min(|t(v)_L|,|t(v)_R|)}\right]\right\} < \infty.
  \label{apr3eq1}\end{equation}
  By (\ref{apr4eq1}), for each $n\geq 1$, there exists $k_n \in\{0,1,\cdots,n-1\}$  such that
  $$\{(i,j):i,j\geq 0, \;\max(i,j)=n-1,\;\sigma(i,j)>0\} \subset \{(k_n,n-1),(n-1,k_n)\}.$$
  Let $(x(n):n\geq 0)$ be the sequence of real numbers such that $x(0)=1$ and
  $$x(n)=x(n-1)+x(k_n),\;\;n\geq 1.$$
  We prove the statement
  \begin{equation}
  |t| = x(d(t)),\;\;t\in \{t^*\}\cup\{t'\in {\cal T}: P_{\sigma}(t')>0\}
  \label{apr17eq5}\end{equation}
  by induction on $|t|$, starting with $|t|=1$. If $|t|=1$, then $t=t^*$ and $|t|=x(d(t))$ is the true statement $1=x(0)$. Now fix $u \in {\cal T}$ for which
   $P_{\sigma}(u)>0$ and we assume as our induction hypothesis that $|t|=x(d(t))$ holds for every $t\in\{t^*\}\cup\{t'\in{\cal T}:P_{\sigma}(t')>0\}$ for which $|t|<|u|$.
   Note that $(d(u_L),d(u_R))$ belongs to the set $\{(d(u)-1,k_{d(u)}),(k_{d(u)},d(u)-1)\}$. The induction hypothesis holds for both $u_L$ and $u_R$, and so
   $$|u| = |u_L| + |u_R| = x(d(u_L)) + x(d(u_R)) = x(d(u)-1)+x(k_{d(u)}) = x(d(u)),$$
   completing the proof of statement (\ref{apr17eq5}).
   We conclude from (\ref{apr17eq5}) that
  for every $t\in{\cal T}$ for which $P_{\sigma}(t)>0$,
  $$\frac{|t(v)|}{\min(|t(v)_L|,|t(v)_R|)} \in \{x(n)/x(k_n):n\geq 1\},\;\;v\in V^1(t).$$
  By (\ref{apr3eq5}), let $m \in {\mathbb{Z}}^+$ be the supremum on the left side of (\ref{apr3eq5}); then $n-1-k_n \leq m$ for $n\geq 1$.
  Since the sequence $(x(n))$ is nondecreasing, $x(n)/x(n-1) \leq 2$ for $n\geq 1$, and so
  $$\frac{x(n)}{x(k_n)} = \prod_{i=k_n+1}^n\frac{x(i)}{x(i-1)} \leq 2^{n-k_n} \leq 2^{m+1},\;\;n\geq 1.$$
  Thus, the left side of (\ref{apr3eq1}) is at most $2^{m+1}$
  and (\ref{apr3eq1}) holds,
  completing our proof.

  \section{Conclusions}

  We have shown that the grammar-based code $(\phi_e,\phi_d)$ on the set $\cal T$ of binary tree structures defined in this paper is asymptotically optimal
  for any binary tree source satisfying the Domination Property and the Representation Ratio Negligibility Property. In typical cases,
  we have found that the Domination Property is easy to verify for a binary tree source, whereas the RRN Property is more troublesome
  to verify. In a subsequent paper \cite{branching}, we investigate more scenarios in which the RRN Property will hold.
  (The one-dimensional binary trees discussed in Example 6 need to be avoided in a binary tree source model, as well
  as some trees derived from these.)
  In \cite{branching}, we also show that $(\phi_e,\phi_d)$ is universal for some families of binary tree sources induced by branching processes (including families of sources
  which were considered in  \cite{miller} from an entropy point of view but not from a compression point of view).

\section*{Acknowledgements}

Our earlier presentation \cite{isit2013} is a summary of the present work. The research of E.-H. Yang in this work is supported in part by the Natural Sciences and Engineering Research Council of Canada under Grant RGPIN203035-11, and by the Canada Research Chairs Program. J.\ Kieffer's research was supported in part by the
NSF Grant CCF-0830457.

 \end{document}